# Order-Disorder Ferroelectric Transition of Strained SrTiO$_3$

**Salva Salmani-Rezaie, Kaveh Ahadi, William M. Strickland and Susanne Stemmer**[1)]

Materials Department, University of California, Santa Barbara, California 93106-5050, USA

[1)] Corresponding author.  Email: stemmer@mrl.ucsb.edu

**ABSTRACT**

$SrTiO_3$ is an incipient ferroelectric that is believed to exhibit a prototype displacive, soft mode ferroelectric transition when subjected to mechanical stress or alloying. We use high-angle annular dark-field imaging in scanning transmission electron microscopy to reveal local polar regions in the room-temperature, paraelectric phase of strained $SrTiO_3$ films, which undergo a ferroelectric transition at low temperatures. These films contain nanometer-sized domains in which the Ti-columns are displaced. In contrast, these nanodomains are absent in unstrained films, which do not become ferroelectric. The results show that the ferroelectric transition of strained $SrTiO_3$ is an order-disorder transition. We discuss the impact of the results on the nature of the ferroelectric transition of $SrTiO_3$.

The classic Devonshire model [1] of a displacive ferroelectric phase transition forms the basis of the thermodynamic theory of some of the most common ferroelectrics, such as $BaTiO_3$ [2]. It assumes that the free energy in the paraelectric phase has a single minimum at zero polarization, whereas the ferroelectric phase has multiple potential wells corresponding to non-zero polarization values and finite Ti ion displacements. In this picture, a long-range polarization develops spontaneously at the diffusion-less phase transition due to the softening of a low-energy, transverse optical (TO) phonon mode. Despite the widespread use of the displacive model, experimental evidence of local structural distortions in the cubic, paraelectric phase of many ferroelectric perovskites [3-7] has led to proposals of alternative models, based on order-disorder transitions [3, 8] or a combination of soft mode and order-disorder pictures [9, 10]. In general, however, there exists no agreement in the literature about the nature of the displacements (static or dynamic) or the degree of correlations among them, both of which would be key to understanding the origin and the nature of ferroelectric transitions.

$SrTiO_3$ is an incipient ferroelectric [11] in its unstrained, pure bulk form, but easily becomes ferroelectric under small perturbations, such as mechanical stresses or alloying [12-15]. Understanding the nature of the ferroelectric transition of $SrTiO_3$ is important for many reasons. It is thought to be a prototype soft mode (incipient) ferroelectric [11, 16, 17]. As such, deviations from the classical Devonshire picture would likely have implications for other materials. The ferroelectric soft mode has also been suggested to play a role in the superconductivity of $SrTiO_3$ [18-23], whose Cooper pairing mechanism remains elusive despite several decades of research [24, 25]. Recently, it was found that the superconducting transition temperature is doubled in compressively strained $SrTiO_3$ films [26, 27] and $O^{18}$-substituted $SrTiO_3$ crystals [28], for which a ferroelectric transition precedes the superconducting state upon cooling. Such "ferroelectric" superconductors have broken inversion symmetry [27], which, in conjunction with spin-orbit coupling, can give rise to unconventional superconductivity [29-34]. Ferroelectricity can thus potentially be used to tune the nature of the superconducting state of $SrTiO_3$.

Although the displacive model is generally used to describe the ferroelectric transition of $SrTiO_3$, indications of an order-disorder component have also emerged. For instance, signatures of polar clusters above the Curie temperature have been reported for $^{18}O$-enriched $SrTiO_3$ and for $SrTiO_3$ containing impurities [6, 7, 35, 36]. A low-temperature phase containing ordered ferroelectric regions has been suggested to exist even in pure, stress-free $SrTiO_3$ [37, 38]. Recently, an Ising model was found to be the best descriptor of the temperature dependence of optical second harmonic generation (SHG) data near the ferroelectric phase transition of compressively strained $SrTiO_3$ films [27]. Such films also show a residual SHG signal in the paraelectric phase and one possible explanation is the existence of polar regions [27, 39].

Images of polar regions above the Curie temperature of ferroelectric $SrTiO_3$ would provide the most direct evidence of an order-disorder transition. Atomic resolution high-angle annular dark-field imaging in scanning transmission electron microscopy (HAADF-STEM) is a powerful technique that is capable of determining the atomic column positions with picometer precision [40-43] and, consequently, should be able to detect regions that contain Ti columns that are off-centered from their non-polar positions. Here, we use HAADF-STEM to measure static Ti-column displacement vectors at room temperature in the paraelectric phase of compressively strained $SrTiO_3$ films, which undergo a ferroelectric transition below 140 K [26, 27, 44]. Unlike Ca-alloyed $SrTiO_3$, these films contain no large concentrations of alloying elements that could form defect clusters. Furthermore, the results can be directly compared with those of unstrained films grown under identical conditions that have identical defect and impurity concentrations, but do not become ferroelectric.

SrTiO$_3$ thin films were grown by hybrid molecular beam epitaxy (MBE) on (001) LSAT [(LaAlO$_3$)$_{0.3}$(Sr$_2$AlTaO$_6$)$_{0.7}$] and (001) SrTiO$_3$ single crystals, respectively, as described elsewhere [45]. Oxygen and Ti were supplied via the metal-organic precursor, titanium tetra-isopropoxide [45]. The in-plane strain of films grown on LSAT is about -1% as long as the film thickness remains below the critical thickness for strain relaxation of about 180 nm [46]. High-resolution 2θ-ω x-ray diffraction scans of the samples are shown in the Supplementary Material [47]. These confirm an out-of-plane lattice parameter of 3.93 Å for SrTiO$_3$ films on LSAT, corresponding to a fully strained film. Interfaces are atomically abrupt with no extended defects (Fig. 1). As-grown films contain oxygen vacancies, which dope the films with about 4×10$^{18}$ cm$^{-3}$ charge carriers. Doped films grown on LSAT were previously shown to undergo a ferroelectric transition at low temperatures to a polar point group (*4mm*) with the polar axis oriented normal to the film plane [27]. The ferroelectric transition temperature of the film studied here was about 140 K [47]. For comparison, a strain-relaxed and oxygen annealed film grown on LSAT was also investigated.

Cross-section TEM samples were prepared using focused ion beam milling with 2 keV Ga ions. STEM-HAADF was carried out using a Thermo-Scientific Talos G2 200×S/TEM (C$_s$= 1.2 mm) at 200 keV with a semi-convergence angle of 10.5 mrad and a HAADF detector angular range of 48–200 mrad (camera length of 125 mm). To improve the signal-to-noise ratio, 20 images (2048 × 2048 pixels, 2 μsec dwell time) were sequentially recorded and cross-correlated. Sample tilts can result in errors in the determination of column positions [48]. We used position averaged convergent beam diffraction (PACBED) [49] to reduce tilts to less than 1 mrad. Atomic column positions were obtained by iterative fitting to a two-dimensional Gaussian function to obtain picometer precision in the Ti column displacement measurements [42]. The Ti column displacement (polarization) vector was defined as the difference between the center of mass of the four surrounding Sr columns and the actual Ti-O column position obtained by 2D Gaussian fitting [see Fig. 1(d)]. All images were acquired at room temperature, well into the paraelectric phase, far above the ferroelectric phase transition.

Figure 2 shows the polarization vectors for two representative images recorded from unstrained (left column) and strained (right column) $SrTiO_3$ films, respectively. The corresponding HAADF-STEM images and PACBED patterns can be found in the Supplemental Material [47]. The arrows in Figs. 2(a,b) represent the directions of the polarization vectors and their magnitudes, which are also indicated by the color scale. The displacements in the unstrained film [Fig. 2(a)] are very small. As discussed in more detail below, they provide a measure of the experimental error (sample instabilities, image distortions, residual sample tilt) in determining the column positions and the influence of oxygen vacancies on the Ti-O columns. In contrast, the displacements in the strained film [Fig. 2(b)] are considerably larger. The average Ti column displacements relative to the center of the unit cell in these images are 4.2±2.1 pm and 18.1±7.4 pm for the unstrained and strained film, respectively. In addition, the films differ in the local alignments of their polarization vectors. Figures 2(c,d) show their directions overlaid on the images. The color scale indicates the polarization orientation in 30-degree intervals. Note that the color scale indicates only the direction of Ti column displacements and is unconnected to their magnitude. The mostly random colors for the unstrained film [Fig. 2(c)] show that there is no preferential direction in the small Ti-column displacements. Despite the small displacements and their random orientation, there appears to be some degree of correlation between the displacements over a very small length scale of a few neighboring unit cells even in the unstrained film (see also additional images in the Supplementary Information [47]). In contrast, Fig. 2(d) is mostly populated by red color, which shows that the Ti column displacements in the strained SrTiO$_3$ film align along the growth direction in a region spanning many unit cells, forming a nanodomain.

Other areas of the strained film contain domains having different polarization directions (Fig. 3). For example, the Ti columns are displaced mostly along [001] in the image shown in Fig. 3(a), whereas in Fig. 3(c) they are displaced predominantly in-plane along [0$\bar{1}$0]. Figure 3(b) shows a region where a small [0$\bar{1}$0] domain is surrounded by a [001] domain. Additional images of other polar domains and of nonpolar regions in both the unstrained and the strained films are shown in the Supplemental Material [47].

For better statistics and to understand the origins of the apparent small, random displacements in the unstrained sample, we analyzed a very large number of images. Figure 4 displays the results for the $x$ and $z$-components of the polarization vectors [defined in Fig. 1(d)] for unstrained and strained films and for the strain-relaxed, oxygen annealed film on LSAT. More than 10,000 Ti columns were analyzed for each sample. The displacements of the unstrained film on SrTiO$_3$ [Fig. 4(a)] have a Gaussian distribution with a full-width at half maximum (FWHM) of 10 pm centered around zero displacement. In contrast, the strained film [Fig. 4(b)] shows bimodal distributions around non-zero values for each of the polarization components. The bimodal distribution can be more easily discerned for the in-plane ($x$) direction (see distribution on the top of the graph). Experimental error causes apparent canting of the polarization vectors, but some regions appear to have real displacements that (in the two-dimensional projection) are aligned along <011> [see also Fig. 2(b)]. The Ti column displacements of the strain-relaxed, oxygen annealed film are centered around zero, indicating no polar domains in this film, consistent with the absence of a ferroelectric transition. Interestingly, the distribution is narrower (FWHM of 8 pm) than for the unstrained film on SrTiO$_3$. We attribute this to the oxygen annealing, which removes the oxygen vacancies. Oxygen vacancies can cause small relaxations of the neighboring Ti columns immediately surrounding the columns [50, 51]. Our results show that these displacements are random, unlike those of the polar nanodomains in the strained film. While the oxygen vacancies themselves do not produce a detectable change in the contrast of HAADF-STEM images, and their concentration is too low to be present with a high likelihood in a single image, they can be detected in a statistical analysis of a large data set of Ti column displacements.

To summarize, our observations of polar nanodomains at room-temperature in the paraelectric phase of strained SrTiO$_3$, featuring correlated off-centering of Ti ions along the tetragonal axes, provide clear evidence in support of an order-disorder transition. A purely displacive transition would not exhibit Ti displacements in the paraelectric phase. The nanodomains only exist in films that subsequently undergo a ferroelectric transition at low temperatures, ruling out other possible proposed origins, such as impurities [7]. While the polarization is normal to the film plane throughout the entire film below the Curie temperature as a result of the compressive in-plane stress [27, 39, 47], significant fractions of the nanodomains have Ti displacements that are oriented in-plane, and some appear to be displaced/projected along <011>. In the paraelectric phase, the different orientations and small size of the polar domains ensure that the fixed polarization charge in the domains remains compensated. This arrangement of nanodomains also makes the films appear nonpolar (or exhibit only a small remnant polarization) in global measurements, such as second harmonic generation, that average over many domains.

We next discuss the impact of our findings on the nature of the ferroelectric transition. Our observations of polar nanodomains in the paraelectric phase are consistent with a type of order-disorder transition originally proposed by Takahasi [8]. In particular, the fact that the displaced Ti ions form small ordered regions within which the displacements are correlated points to strong electrostatic interactions among the Ti ions already above the ferroelectric Curie temperature. This has several implications with regard to the nature of the phase transition. At the transition to the ferroelectric phase, in which the ferroelectric polarization is oriented globally out-of-plane, these existing dipoles must reorient, causing nanodomains to grow and percolate to form the long-range ordered state. The nature of the transition is therefore from a locally ordered, globally random phase to a globally ordered phase. The fact that the nanodomains exist above the transition temperature shows that the transition is driven by the strong electrostatic interactions of the Ti ions. Moreover, signatures of small, short-range correlated, but randomly oriented, displacements in the unstrained film point to interactions that exist to very high temperatures (relative to the phase transition temperature) and supports the possibility of earlier suggestions [37, 38] of low temperature nanodomains even in unstrained, pure $SrTiO_3$. In other words, above the Curie temperature and with decreasing temperature, the interactions among neighboring unit cells become increasingly strong, causing the nanodomains to grow and the off-centering to develop along preferred orientations within a globally unpolarized, high-symmetry phase. The nanodomains are therefore an essential ingredient for the transition to take place.

It is also important to note that because of the long exposure time, HAADF-STEM images cannot detect dynamic polar fluctuations. Thus, our observations show static displacements within polar regions in the paraelectric phases. A dynamically disordered phase, which forms the basis for some of the theoretical proposals for $BaTiO_3$, would therefore be difficult to reconcile with our findings. We note the similarities with recent reports for $BaTiO_3$, which show static polar nanodomains [52, 53] as well as static displacements in the cubic phase [54], and PbTe, which is close to a ferroelectric instability [55]. These similarities suggest that an order-disorder transition that is characterized by static, locally ordered dipoles in the paraelectric phase that transition to a long range ordered phase might be a more common feature than previously thought. The results have important consequences for modeling the ferroelectric transition of $SrTiO_3$, most of which were based on a displacive transition. Moreover, theories that link ferroelectricity to the superconductivity of $SrTiO_3$, especially those that connect the superconducting pairing mechanisms in $SrTiO_3$ to the soft mode behavior, should consider the complexity of the observed ferroelectric transition.

**Acknowledgments**

This work was supported by the U.S. Department of Energy (Grant No. DEFG02-02ER45994). Film growth experiments were supported by a MURI program of the Army Research Office (Grant No. W911NF-16-1-0361). This work made use of the MRL Shared Experimental Facilities, which are supported by the MRSEC Program of the US National Science Foundation under Award No. DMR 1720256.

# Figure Captions

**Figure 1:** Cross-section HAADF-STEM images of SrTiO$_3$ films grown on (a,c) LSAT and (b) SrTiO$_3$. A schematic illustrating the polarization vector and its components is shown in (d).

**Figure 2:** Polarization vectors for SrTiO$_3$ films grown on (a,c) SrTiO$_3$ and (b,d) LSAT. The arrows in the figures in the top row indicate the magnitude (also indicated by the color scale) and orientation of the polarization vectors. The images in the bottom row display the direction of the polarization vectors overlaid on the HAADF image, with each color corresponding to a 30 degree interval of the polarization directions.

**Figure 3:** Additional regions of the film on LSAT showing nanodomains of different polarization orientations. The colors indicate the direction of the polarization vectors, with each color corresponding to a 30-degree interval of the polarization directions. The colored dots are overlaid on the actual HAADF-STEM images. The corresponding polarization vectors are shown in the Supplementary Information [47].

**Figure 4:** Magnitude of displacement vector components [defined in Fig. 1(d)] acquired from multiple regions of films on (a) SrTiO$_3$ (b) LSAT (c) on LSAT after oxygen annealing.

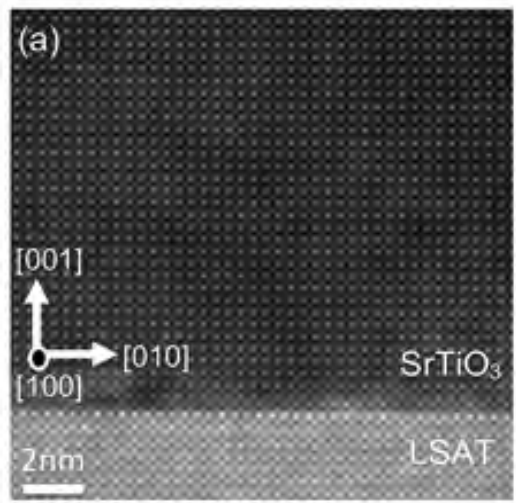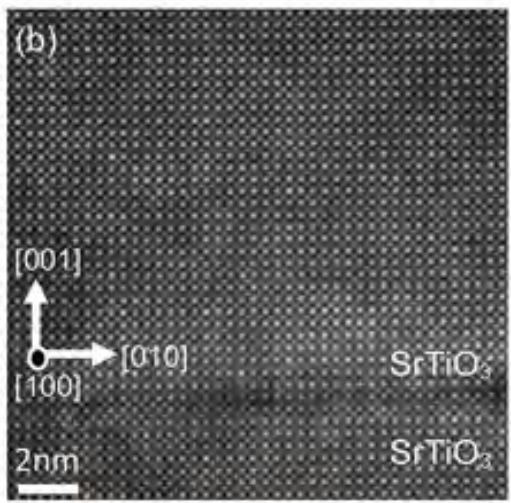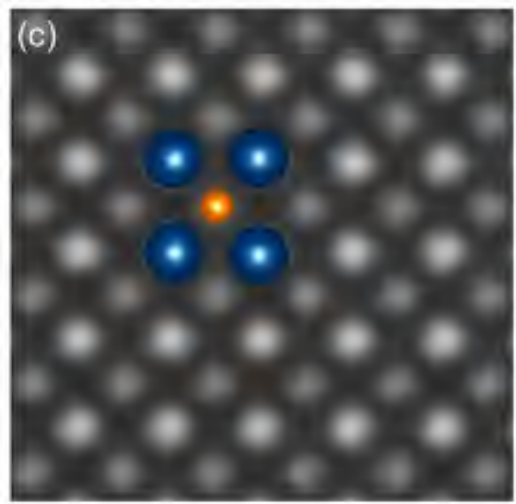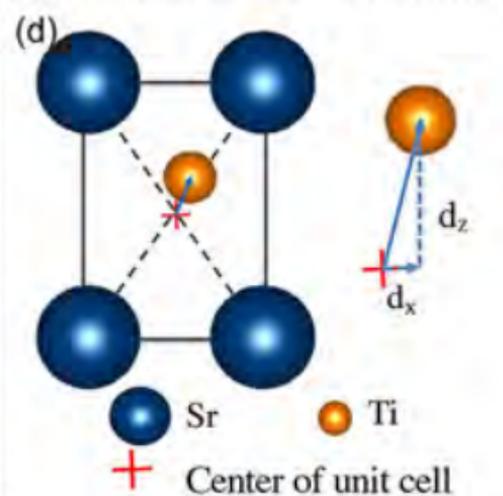

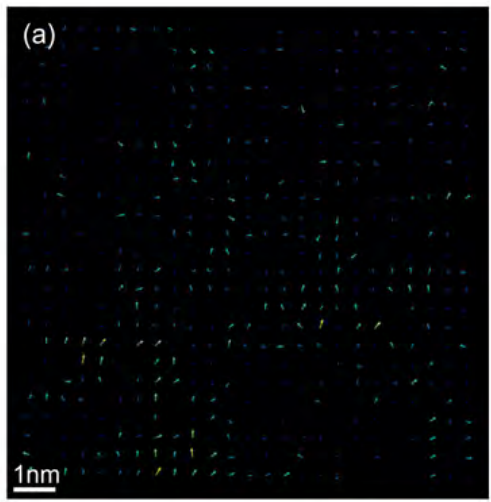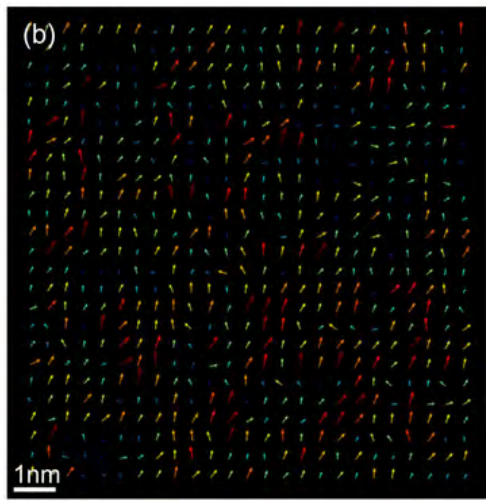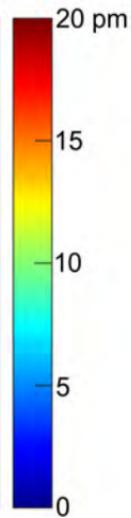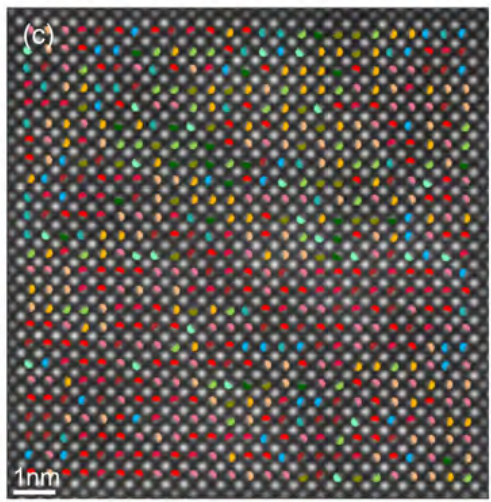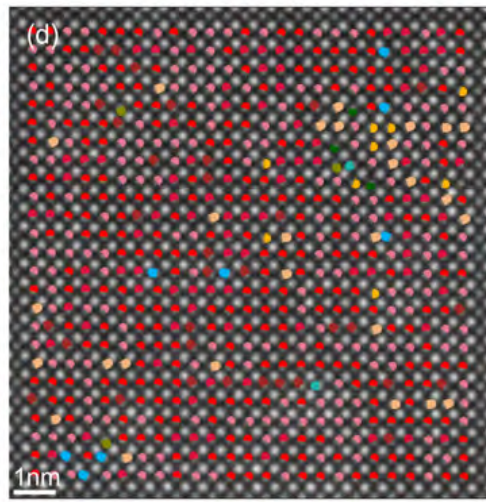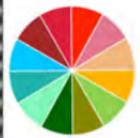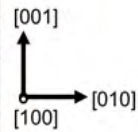

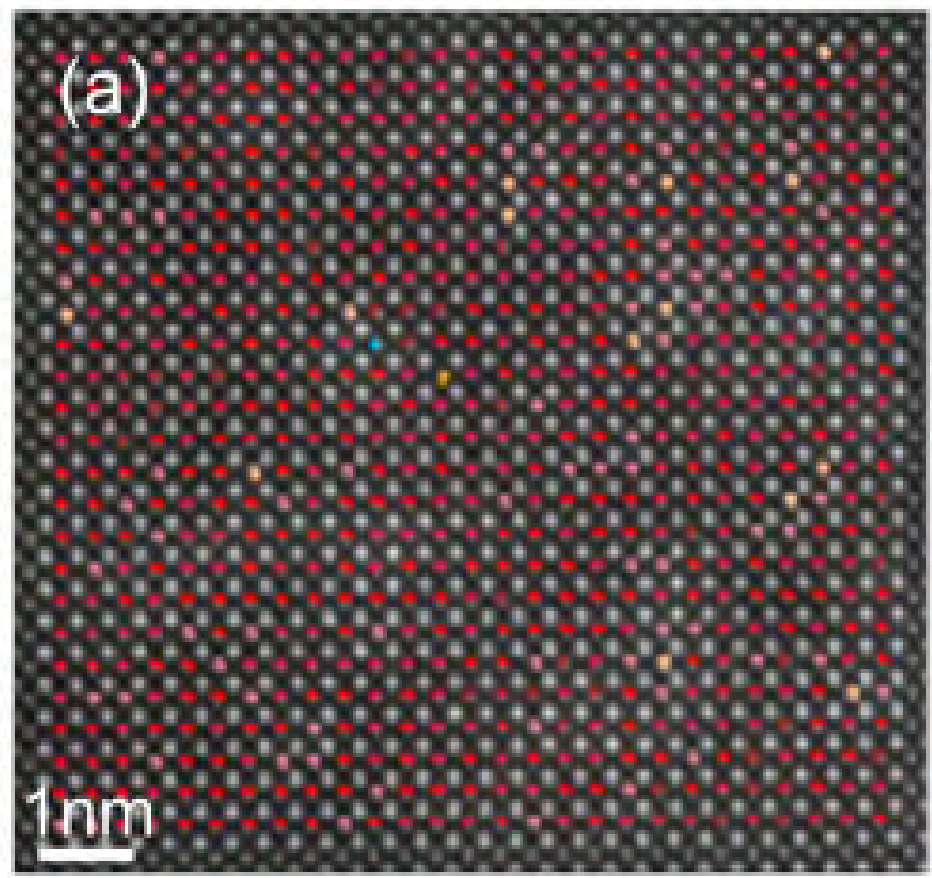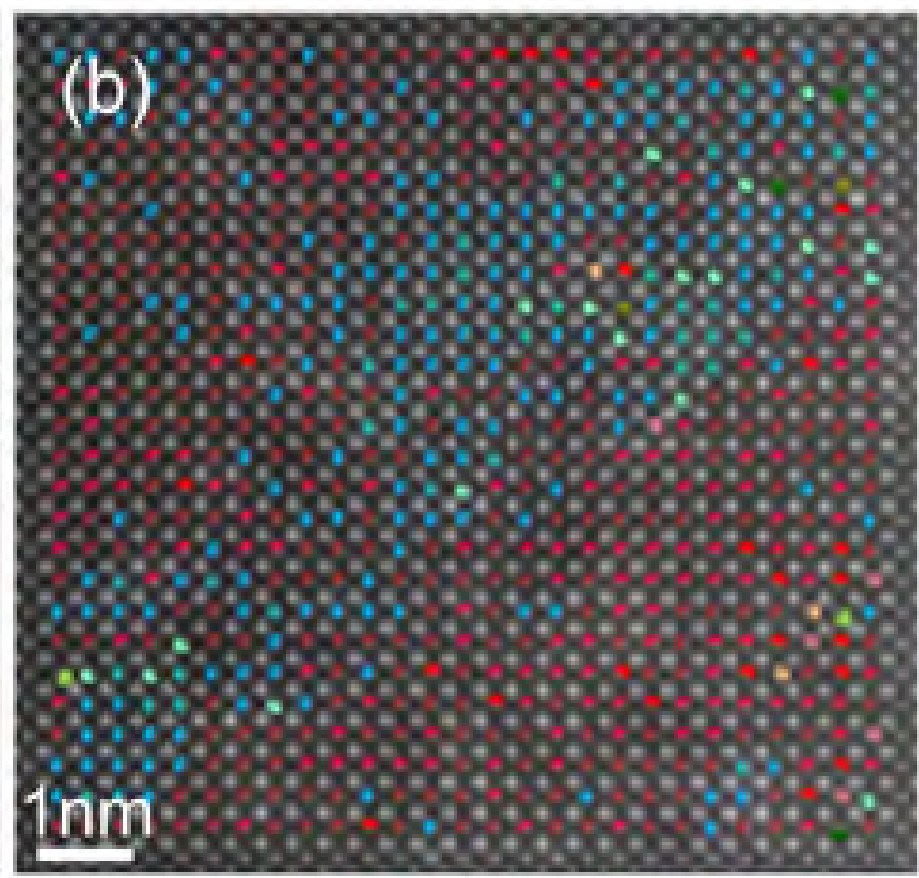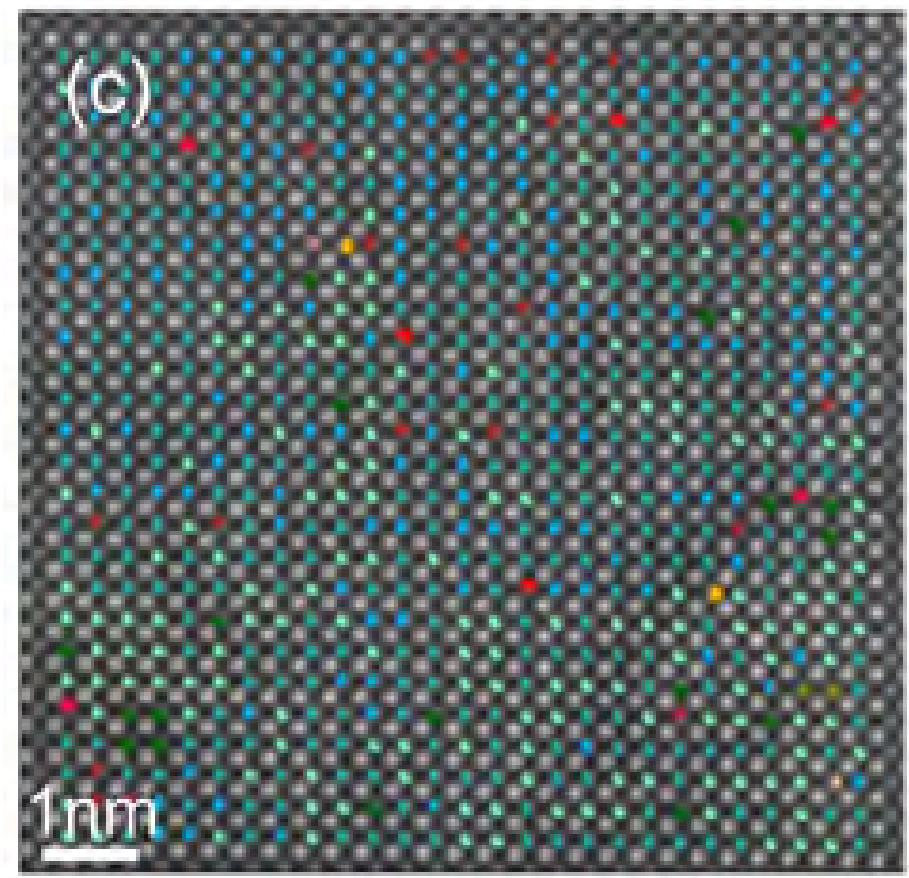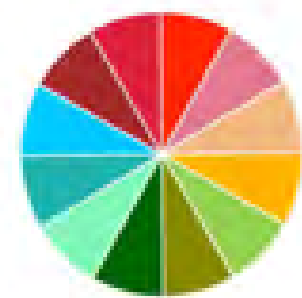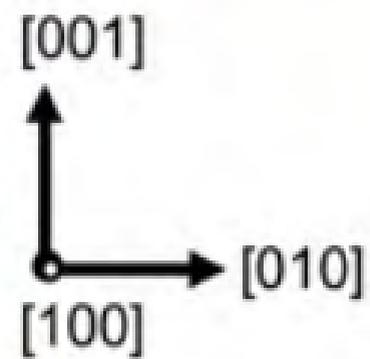

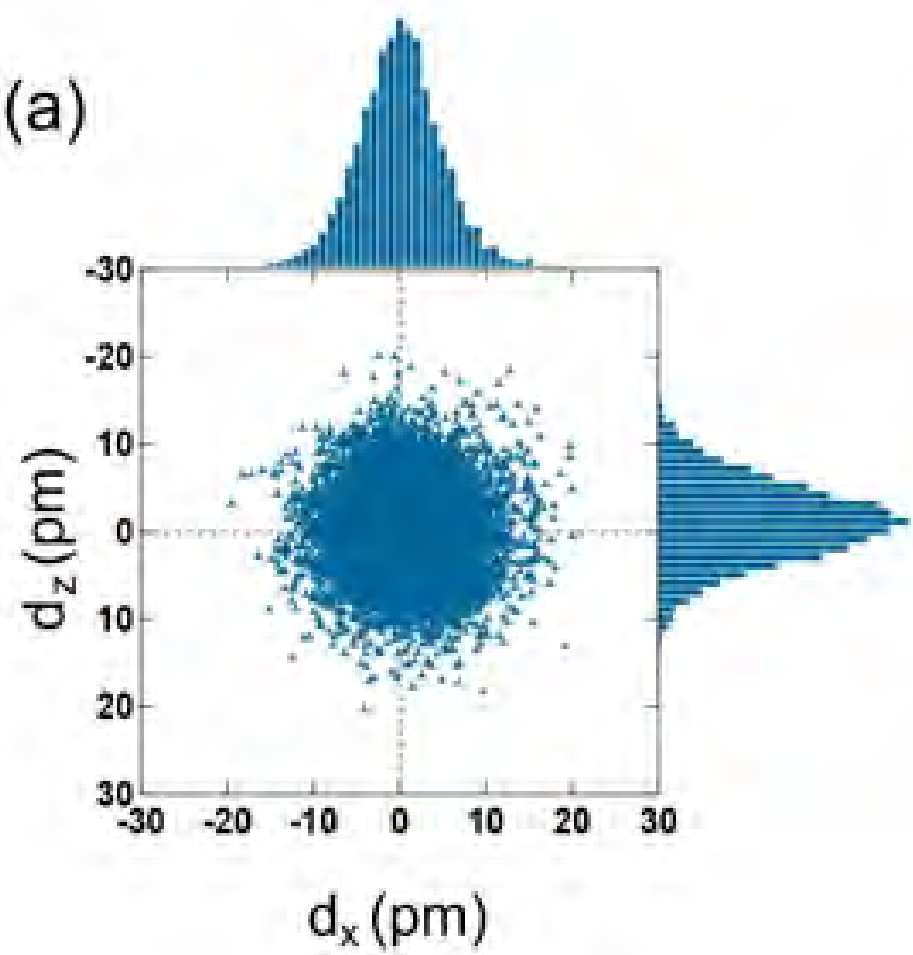 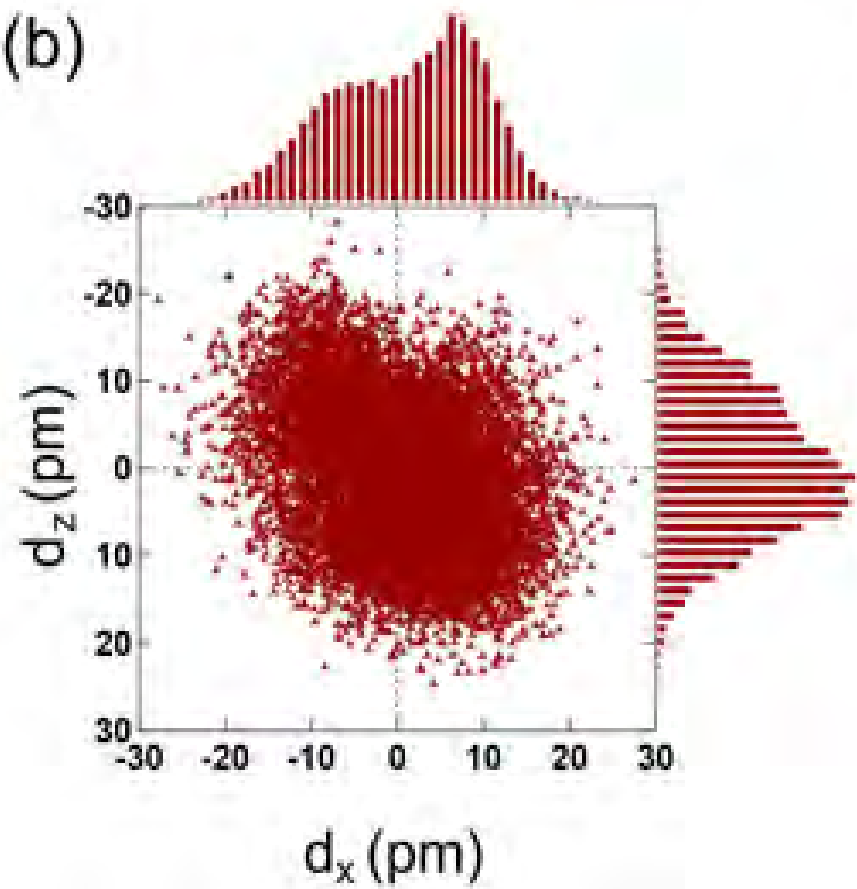 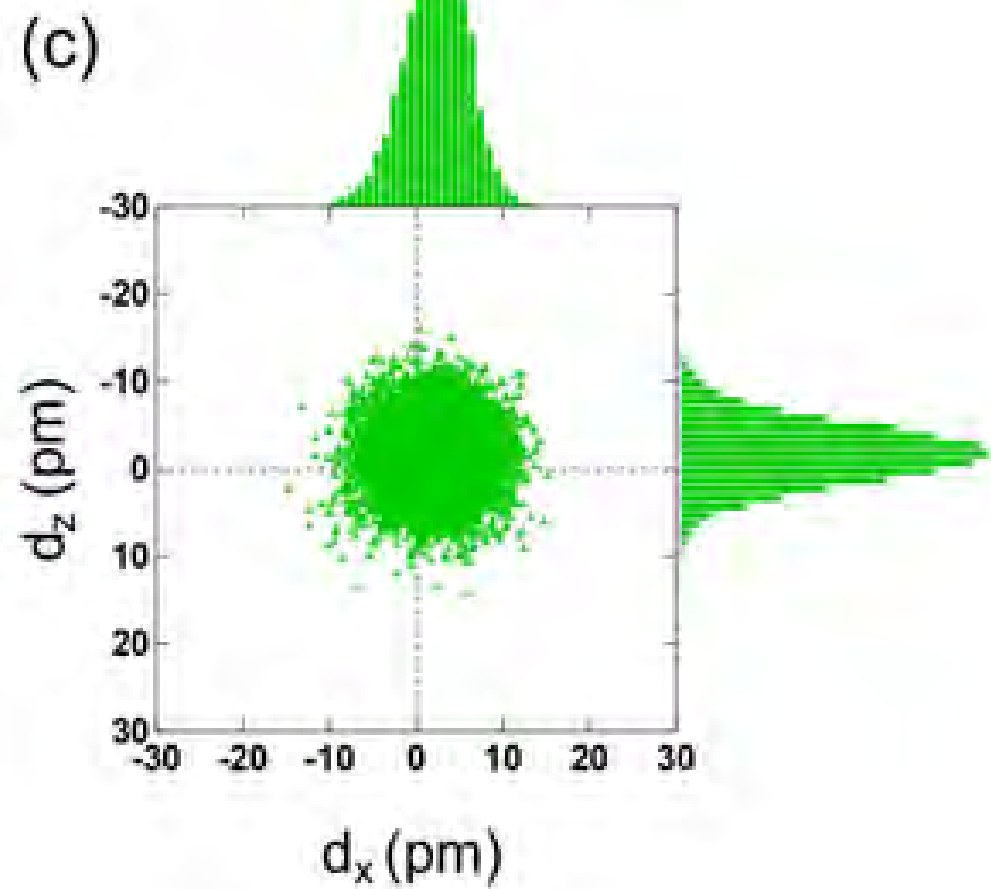

# Supplementary Information
# Order-Disorder Ferroelectric Transition of Strained SrTiO$_3$
### Salva Salmani-Rezaie, Kaveh Ahadi, William M. Strickland and Susanne Stemmer

## X-ray diffraction

A Philips Panalytical X'Pert thin-film diffractometer with Cu K$_\alpha$ radiation was used for high-resolution x-ray diffraction (XRD) characterization of the films. Figure S1 shows 2θ-ω scans around the 002 reflections of the SrTiO$_3$ films for the strained film on LSAT, the unstrained film grown on SrTiO$_3$, and the oxygen annealed, strain-relaxed film on LSAT. Both the strained film and the unstrained film on SrTiO$_3$ exhibit thickness fringes indicating coherency. The out-of-plane lattice parameter of the film in LSAT is 3.930±0.001 Å, as expected for a fully strained SrTiO$_3$ film on LSAT. The annealed SrTiO$_3$ film grown on LSAT substrate is relaxed. The loss of coherency is evident not only from the reduced out-of-plane lattice parameter, but also in the absence of thickness fringes.

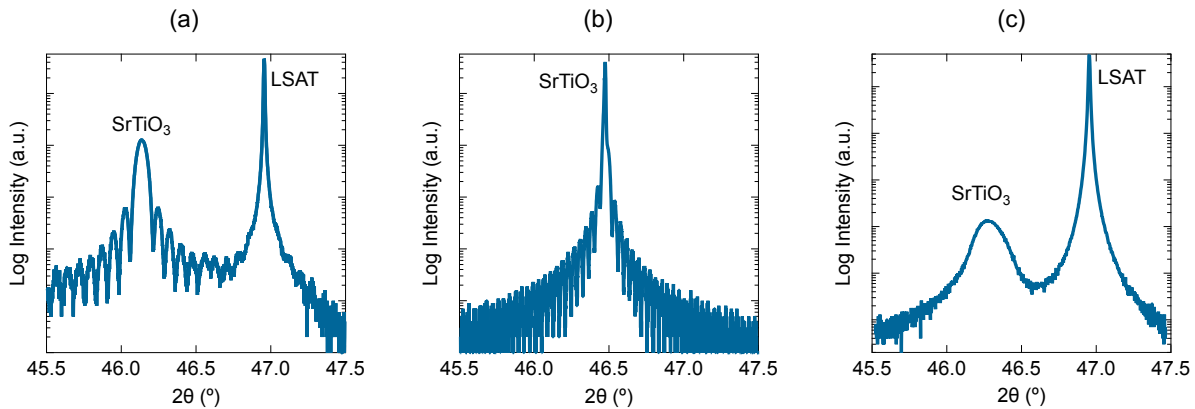

**Figure S1:** 2θ-ω XRD scans near the 002 SrTiO$_3$ film reflections (a) on LSAT (b) on SrTiO$_3$ and (c) annealed film on LSAT.

## Transport Properties

Figure S2 shows the temperature dependence of the sheet resistance ($R_s$) of the strained SrTiO$_3$ film on LSAT, measured between 300 and 1.7 K. Oxygen vacancies dope the film, resulting in a Hall carrier density of 4×10$^{18}$ cm$^{-3}$ at room temperature. The sample shows metallic behavior at high temperatures with a resistance anomaly occurring at ~140 K. This anomaly (a small upturn in resistance) has previously been shown to be an indication of the ferroelectric transition. A large increase in the second harmonic generation signal occurs at the resistance anomaly [1]. The temperature of the transition is similar to previous reports for SrTiO$_3$ on LSAT [2, 3].

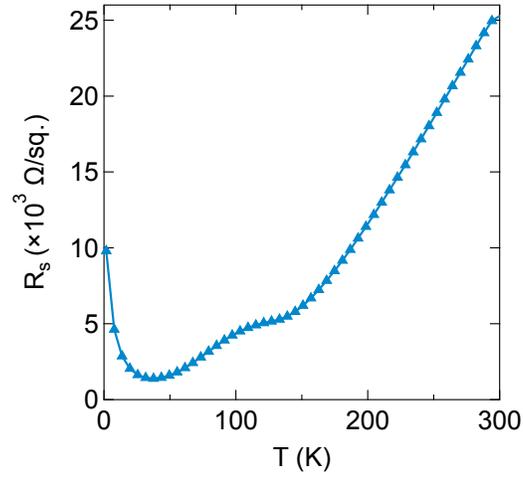

**Figure S2:** $R_s$ as a function of temperature for the strained SrTiO$_3$ film grown on LSAT.

## PACBED and HAADF-STEM images

PACBED patterns were recorded for all analyzed images to ensure that the sample tilt was less than 1 mrad. Figure S3 shows the PACBED patterns and the corresponding HAADF-STEM images for the data shown in Fig. 2 in the main text.

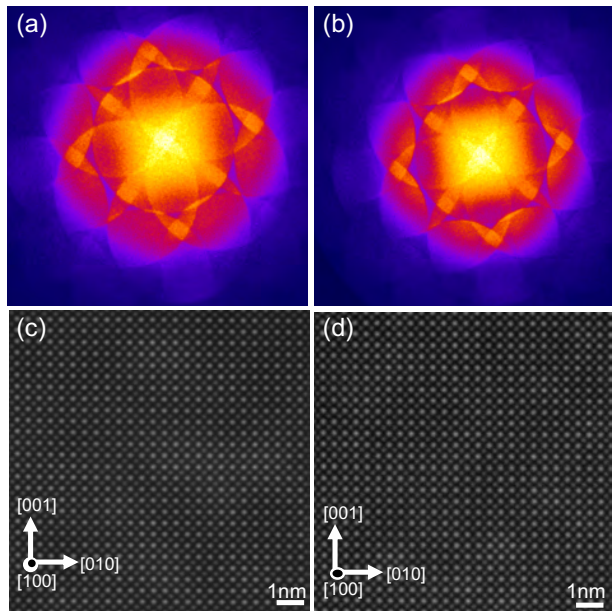

**Figure S3:** Experimental PACBED patterns for a region of (a) the unstrained SrTiO$_3$ film grown on SrTiO$_3$ and (b) strained SrTiO$_3$ films grown on LSAT. The corresponding HAADF-STEM images are shown in (c) and (d).

## Polarization vectors

Figure S4 shows the polarization vectors of the images analyzed in Fig. 3 of the main text.

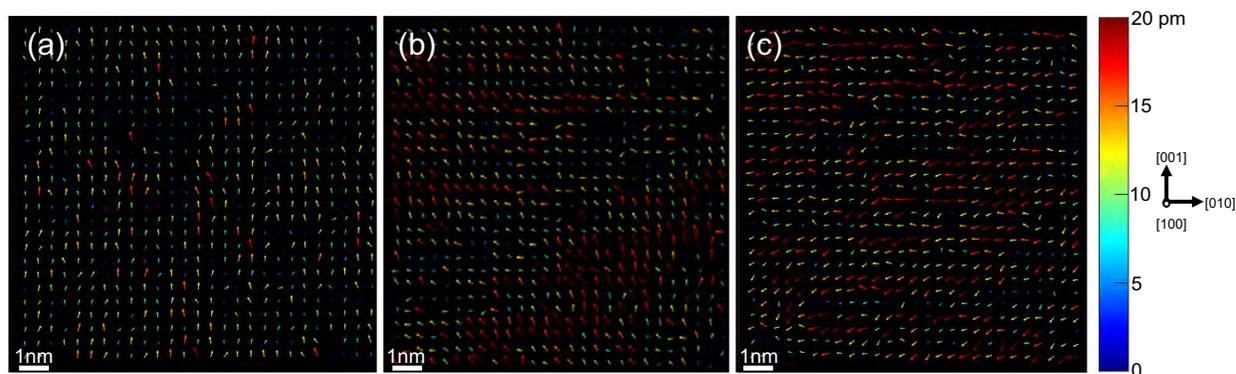

## Additional polarization vector measurements

Figure S5 shows additional measurements of polarization vectors for the strained SrTiO$_3$ film, obtained images from different regions of the sample. These images confirm the existence of nanodomains throughout and show their varying sizes.

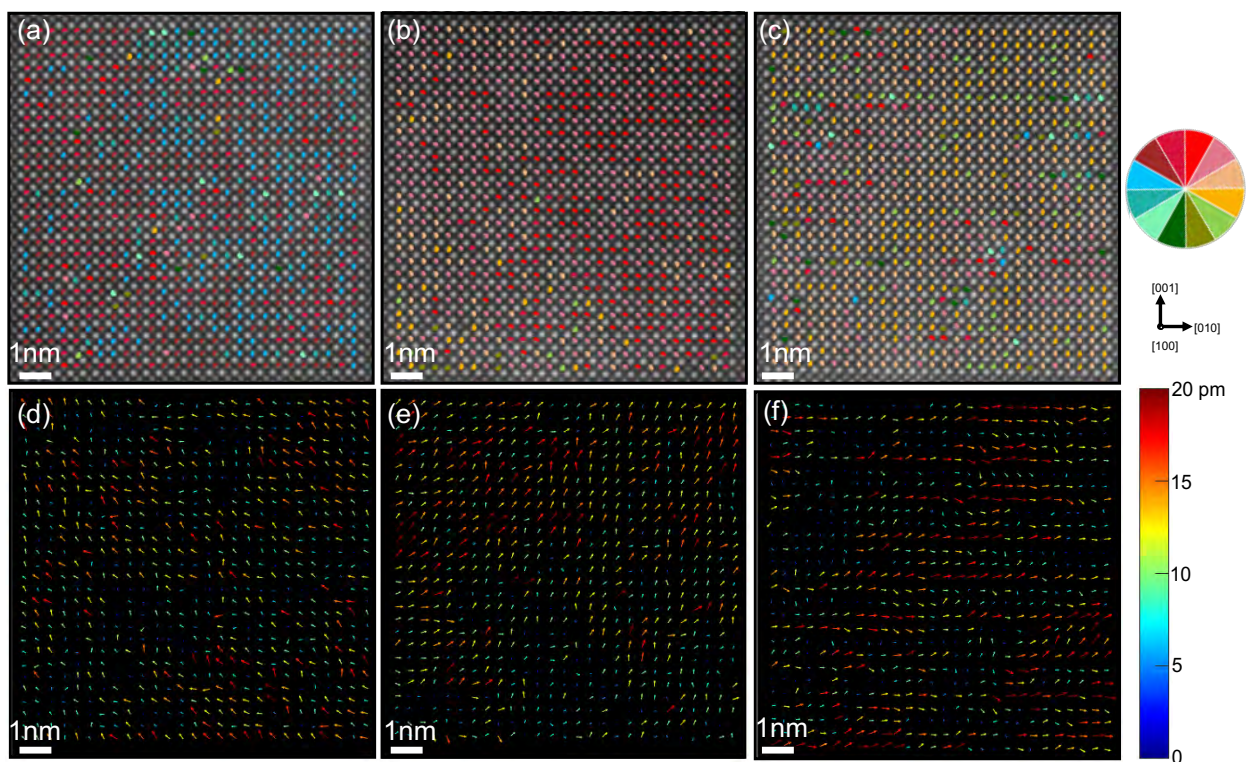

directions of [001] and [010] (a,d), [001] and [010] (b,e), and [010] (c,f) are present.

Figure S6 shows additional polarization maps for the strained SrTiO$_3$ film, in this case for regions that contain apparent non-polar regions. In these regions, the Ti atom off-centering appears random. These regions are larger than a single unit-cell. Due to the projection issue, it is possible, however, that these regions contain correlated Ti-off-centering parallel to the beam.

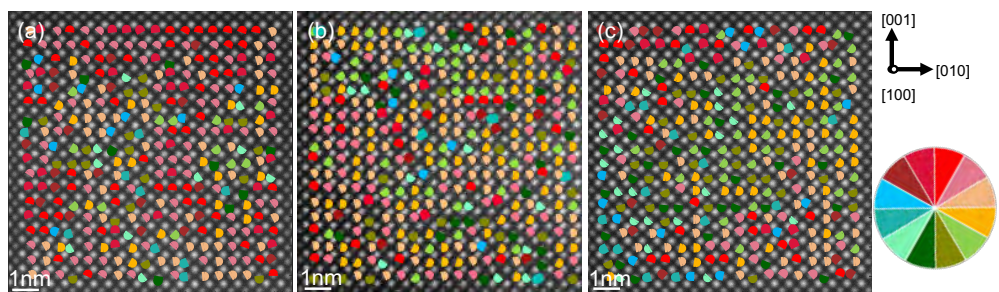
**Figure S6:** Polarization maps containing apparent non-polar regions in the strained SrTiO$_3$ film.

Figure S7 shows additional polarization maps for the unstrained SrTiO$_3$ film from different regions of the sample. Compared to the paraelectric phase of strained SrTiO$_3$, the Ti atom displacements are clearly more random. However, even here, on a much smaller length scale, correlations between the lattice distortions can sometimes be observed.

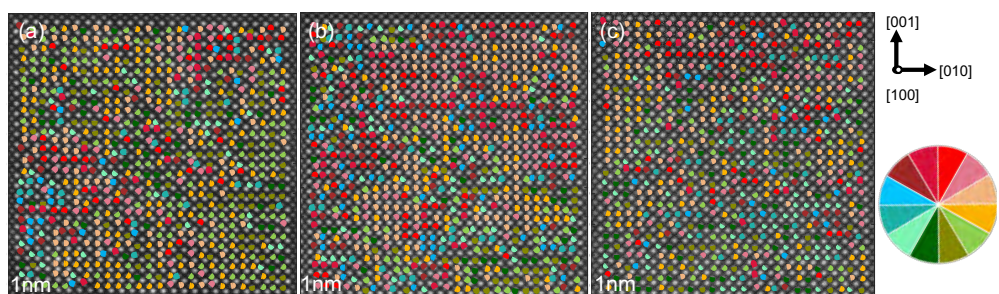
**Figure S7**: Additional polarization maps of unstrained SrTiO$_3$ film.

## Low-temperature phase

A liquid nitrogen-cooled sample holder was used to acquire images at low temperatures (110 K). Unfortunately, HAADF-STEM measurements below room temperatures suffer from thermal drift and vibrations caused by liquid nitrogen coolant, which makes obtaining larger data sets very challenging. Figure S8 shows a polarization map and polar vectors acquired at 110 K. The polar axis is oriented normal to the film plane, as expected from the literature data discussed in the main text.

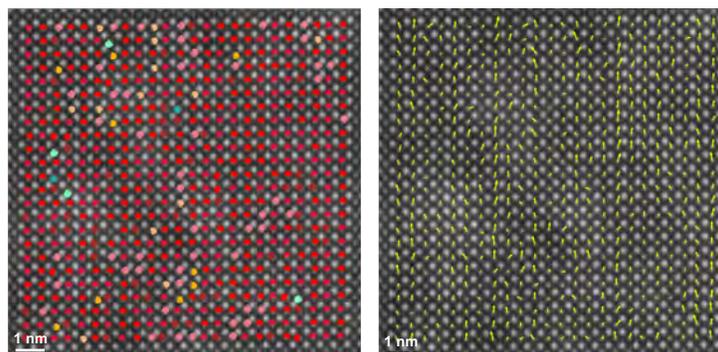
**Figure S8**: Polarization map of strained SrTiO$_3$ film recorded at 110 K.